\documentclass[11pt,a4paper]{article}
\usepackage{jheppubm}
\usepackage{mathrsfs}
\usepackage{csquotes}
\usepackage{amsthm}
\newtheorem{theorem}{Theorem}

\title{(Reply to)$^{\textbf{2}}$ ``Comment on `Flavor invariants and renormalization-group equations in the leptonic sector with massive Majorana neutrinos' ''}

\author{Jianlong Lu}

\affiliation{Department of Physics, National University of Singapore,\\
2 Science Drive 3, 117551, Singapore}

\emailAdd{jianlong\_lu@u.nus.edu}

\abstract{We respectfully reply to Wang et al.'s reply to our comment to their [\emph{JHEP} \textbf{09} (2021) 053]. Some more subtleties and details are discussed for hopefully better clarification and understanding, especially about the conditions in Hilbert's finiteness theorem. We also disprove Wang et al.'s incorrect assertion in their reply [arXiv:2110.13865 [hep-ph]] that there is an isomorphism between ${\rm U}(n,\mathbb{C})$ and ${\rm GL}(n,\mathbb{C})$. We give two proofs for our argument. One proof is in the context of Lie group for arbitrary positive integer $n$, and the other proof is in the context of abstract group for $n=3$ which can also be generalized to $n\geq 3$. In order to better illustrate the intrinsic differences between a compact Lie group and its complexification, we choose  ${\rm SU}(n,\mathbb{C})$ and ${\rm SL}(n,\mathbb{C})$ as a pair of examples and give two proofs for the generally non-isomorphic relation between them. One proof is in the context of Lie group for integers $n\geq 2$, and the other proof is in the context of abstract group for $n=3$ which can again be generalized to $n\geq 3$.}  

\arxivnumber{}

\begin{document}
\maketitle
\flushbottom

\section{Introduction}
\label{s1}
This series of discussions \cite{wyz,lu,wyz2} originated from Wang et al.'s effort in \cite{wyz} to construct a finite generating set for the ring of unitary invariants of the Hermitian matrices $\{H_{l},H_{\nu},G_{l\nu}^{(k)} \}$ with $k\in\mathbb{N}^{+}$ less than the number of generations of leptons. The disputed point is whether the unitary group ${\rm U}(n,\mathbb{C})$ is a reductive group so that this can be directly applied to Hilbert's finiteness theorem to prove that the above ring of unitary invariants of the Hermitian matrices $\{H_{l},H_{\nu},G_{l\nu}^{(k)} \}$ is finitely generated over $\mathbb{C}$. Their answer is affirmative, but ours is not.\\
The remaining part of this manuscript is organized as follows. In section \ref{s2}, we briefly clarify our discussion in \cite{lu} about the scope of  flavor invariants of a given set of matrices under certain transformation (in the context of our discussion it is simply $\{H_{l},H_{\nu},G_{l\nu}^{(k)} \}$ under ${\rm U}(n,\mathbb{C})$). In section \ref{s3}, which is the core of this manuscript, we detailedly discuss the relation among reductive groups, compact Lie groups and Hilbert's finiteness theorem. In particular, we point out an incorrect assertion presented in \cite{wyz2} about the group-theoretic relation between ${\rm U}(n,\mathbb{C})$ and ${\rm GL}(n,\mathbb{C})$, about which we also give two proofs for our argument. One proof is in the context of Lie group for arbitrary positive integer $n$, and the other proof is in the context of abstract group for $n=3$ which can also be generalized to $n\geq 3$. Here by ``in the context of abstract group'', we mean all extra structures equipped on groups, such as topological and geometrical properties, are ignored and only the purely group-theoretic properties are considered in the analysis. Another compact Lie group ${\rm SU}(n,\mathbb{C})$ and its complexification ${\rm SL}(n,\mathbb{C})$ are chosen for better illustration of the intrinsic differences between a compact Lie group and its complexification. We prove that in the context of Lie group ${\rm SU}(n,\mathbb{C})$ and ${\rm SL}(n,\mathbb{C})$ are not isomorphic for integers $n\geq 2$. We also give a proof for that in the context of abstract group ${\rm SU}(n,\mathbb{C})$ and ${\rm SL}(n,\mathbb{C})$ are not isomorphic for $n= 3$, which can be easily generalized to cases with integers $n\geq 3$.

\section{About the scope of flavor invariants of a given set of square matrices under certain transformation}
\label{s2}
Trivial thing comes first. This section corresponds to section 5 in our \cite{lu}, which is about the scope of flavor invariants of a given set of square matrices under certain transformation (in the context of our discussion it is $\{H_{l},H_{\nu},G_{l\nu}^{(k)} \}$ under ${\rm U}(n,\mathbb{C})$).\\
The relation between the polynomial invariants of a given set of Hermitian matrices under certain transformation and other quantities invariant under the same transformation is very easy to understand. In the case of algebraically closed fields with characteristic $0$ such as $\mathbb{C}$, it has been pointed out as first fundamental theorem in \cite{procesi} and \cite{procesi2} that the ring consisting of the former under ${\rm GL}(n,\mathbb{C})$ is generated by the elements ${\rm Tr}(M)$ with $M$ being any monomial in the given set of Hermitian matrices. And we know that the ring consisting of the former under ${\rm U}(n,\mathbb{C})$ is also generated by the elements ${\rm Tr}(M)$ with $M$ being any monomial in the given set of Hermitian matrices, due to the isomorphism between the ring of ${\rm U}(n,\mathbb{C})$ invariants of $j$ complex square matrices $\{x_{1},...,x_{j}\}$ and the ring of ${\rm GL}(n,\mathbb{C})$ invariants of $2j$ matrices $\{x_{1},...,x_{j},x_{1}^{\dagger},...,x_{j}^{\dagger}\}$. Then from the nontrivial mass dimensions of elements in $\{H_{l},H_{\nu},G_{l\nu}^{(k)} \}$ and the dimensionless nature of Jarlskog invariant, one can immediately see that Jarlskog invariant is not a flavor invariants of $\{H_{l},H_{\nu},G_{l\nu}^{(k)} \}$ under unitary transformation and cannot be expressed as a polynomial of the basic invariants, although Jarlskog invariant is indeed unchanged under such unitary transformation. \textbf{Both of Wang et al. and us agree on this statement, according to \cite{wyz2} and \cite{lu}.} This issue is quite trivial and easy for anyone to understand. The reason we mentioned this issue in \cite{lu} (and are mentioning it again in this manuscript) is that the usage of relevant terminology in \cite{wyz} is sloppy and thus somewhat misleading for other readers who are not familiar with invariant theory of square matrices. In section 1 of \cite{wyz}, Wang et al. themselves say ``\emph{The first flavor invariant has been constructed by Jarlskog in refs. \cite{jar1,jar2} in order to characterize the CP violation in the quark sector.}'' But in the abstract of \cite{wyz}, Wang et al. themselves say ``\emph{Any flavor invariants can be expressed as the polynomials of those 34 basic invariants in the generating set.}'' Similar statements can also be found for example in section 1 of \cite{wyz} by Wang et al. themselves: ``\emph{According to the classical invariant theory \cite{stur,derksen}, the other flavor invariants can be expressed as the polynomials of the basic ones.}'' \textbf{I believe any careful reader can see the inadequacy in the usage of terminology in \cite{wyz} at this issue.} Hopefully our clarification in \cite{lu} and this manuscript can reduce the possibility of confusion for interested readers who are not familiar with invariant theory of square matrices.

\section{About reductive groups, compact Lie groups and Hilbert's finiteness theorem}
\label{s3}

\subsection{The relation between ${\rm U}(n,\mathbb{C})$ and ${\rm GL}(n,\mathbb{C})$}
In section 2 of \cite{wyz2}, Wang et al. themselves assert that ``\emph{Although ${\rm U}(N,\mathbb{C})$ is not an algebraic group by definition, there is an isomorphism between the compact Lie group ${\rm U}(N,\mathbb{C})$ to the general linear group ${\rm GL}(N,\mathbb{C})$, which is a reductive algebraic group.}'' We respectfully disagree with this assertion. In the context of Lie group, it is well-known that the unitary group ${\rm U}(n,\mathbb{C})$ and the general linear group ${\rm GL}(n,\mathbb{C})$ are not isomorphic. Here is one proof.
\begin{proof}
Suppose ${\rm U}(n,\mathbb{C})$ and ${\rm GL}(n,\mathbb{C})$ are isomorphic, then by definition there must exist a Lie group isomorphism between them. We know that a Lie group isomorphism is also a diffeomorphism \cite{jlee}, which preserves topological properties including compactness. However, ${\rm U}(n,\mathbb{C})$ is compact but ${\rm GL}(n,\mathbb{C})$ is not compact. Therefore such Lie group isomorphism does not exist.
\end{proof}
\noindent Furthermore, even in the context of abstract group, in which the extra structures equipped on groups such as topology are all ignored, one can still prove that ${\rm U}(n,\mathbb{C})$ and ${\rm GL}(n,\mathbb{C})$ are not isomorphic for $n\geq 3$. Here is one proof for the case $n=3$, which is most relevant to our discussion. We have made the proof as detailed as possible (unnecessarily detailed for most mathematicians) so that anyone with basic knowledge of group theory should have no difficulty in understanding it.
\begin{proof}
Suppose ${\rm U}(3,\mathbb{C})$ and ${\rm GL}(3,\mathbb{C})$ are isomorphic, with one isomorphism denoted by $\rho:{\rm GL}(3,\mathbb{C}) \rightarrow {\rm U}(3,\mathbb{C})$. Then for any subset $S$ of ${\rm GL}(3,\mathbb{C})$, the centralizer of subset $S$ of ${\rm GL}(3,\mathbb{C})$ is isomorphic to the centralizer of subset $\rho(S)$ of ${\rm U}(3,\mathbb{C})$. We choose $S = \{ \begin{pmatrix} 1 & 1 & 0\\ 0 & 1 & 1\\ 0 & 0 & 1\end{pmatrix}\}\subset {\rm GL}(3,\mathbb{C})$. It is obvious that $\begin{pmatrix} 1 & 1 & 0\\ 0 & 1 & 1\\ 0 & 0 & 1\end{pmatrix}$ is a unipotent element of ${\rm GL}(3,\mathbb{C})$ since all of its eigenvalues are equal to $1$. Then the centralizer of the subset $S$ of ${\rm GL}(3,\mathbb{C})$ is the subgroup of ${\rm GL}(3,\mathbb{C})$ with elements $\begin{pmatrix} a_{1} & a_{2} & a_{3}\\ b_{1} & b_{2} & b_{3}\\ c_{1} & c_{2} & c_{3}\end{pmatrix}$ satisfying $\begin{pmatrix} a_{1} & a_{2} & a_{3}\\ b_{1} & b_{2} & b_{3}\\ c_{1} & c_{2} & c_{3}\end{pmatrix} \begin{pmatrix} 1 & 1 & 0\\ 0 & 1 & 1\\ 0 & 0 & 1\end{pmatrix} = \begin{pmatrix} 1 & 1 & 0\\ 0 & 1 & 1\\ 0 & 0 & 1\end{pmatrix} \begin{pmatrix} a_{1} & a_{2} & a_{3}\\ b_{1} & b_{2} & b_{3}\\ c_{1} & c_{2} & c_{3}\end{pmatrix}$. It is easy to obtain $b_{1}=c_{1}=c_{2}=0$, $a_{1} = b_{2} = c_{3}$ and $a_{2} = b_{3}$. Therefore the centralizer of the subset $S$ of ${\rm GL}(3,\mathbb{C})$ is $\{ \begin{pmatrix} a_{1} & a_{2} & a_{3}\\ 0 & a_{1} & a_{2}\\ 0 & 0 & a_{1}\end{pmatrix}: a_{1},a_{2},a_{3}\in\mathbb{C}, a_{1}\neq 0 \}$. Since the center of ${\rm GL}(3,\mathbb{C})$ is $\{\begin{pmatrix} \lambda & 0 & 0\\ 0 & \lambda & 0\\ 0 & 0 & \lambda\end{pmatrix}: \lambda\in \mathbb{C}^{*} \}$, one can see that the centralizer of the subset $S$ of ${\rm GL}(3,\mathbb{C})$ quotient the center of ${\rm GL}(3,\mathbb{C})$ is simply $\{ \begin{pmatrix} 1 & a_{2} & a_{3}\\ 0 & 1 & a_{2}\\ 0 & 0 & 1\end{pmatrix}: a_{2},a_{3}\in\mathbb{C} \}$. For a given pair $(a_{2},a_{3})\in\mathbb{C}^{2}$, suppose there exists a positive integer $k$ such that $\begin{pmatrix} 1 & a_{2} & a_{3}\\ 0 & 1 & a_{2}\\ 0 & 0 & 1\end{pmatrix}^{k} =  \begin{pmatrix} 1 & 0 & 0\\ 0 & 1 & 0\\ 0 & 0 & 1\end{pmatrix}$. The information retrieved from the $(1,2)$-entry or the $(2,3)$-entry tells us that $a_{2}   = 0$. With $a_{2} = 0$ and the information retrieved from the $(1,3)$-entry, we have  $a_{3} = 0$. Hence the only elements in the centralizer of the subset $S$ of ${\rm GL}(3,\mathbb{C})$ quotient the center of ${\rm GL}(3,\mathbb{C})$ with finite order is the identity matrix. Therefore by definition the centralizer of the subset $S$ of ${\rm GL}(3,\mathbb{C})$ quotient the center of ${\rm GL}(3,\mathbb{C})$ is torsion-free.\\
Next we focus on ${\rm U}(3,\mathbb{C})$. The map $\rho$ takes $\begin{pmatrix} 1 & 1 & 0\\ 0 & 1 & 1\\ 0 & 0 & 1\end{pmatrix}$ to some element in ${\rm U}(3,\mathbb{C})$, named $u$ for convenience. We only know some partial information of this element in ${\rm U}(3,\mathbb{C})$, such as its inifnite order, but not the concrete form. However, we know that any element in ${\rm U}(3,\mathbb{C})$ is conjugate to some element in the Cartan subgroup $T=\{\begin{pmatrix} e^{i\theta_{1}} & 0 & 0\\ 0 & e^{i\theta_{2}} & 0\\ 0 & 0 & e^{i\theta_{3}}\end{pmatrix}: \theta_{1},\theta_{2},\theta_{3}\in\mathbb{R}\}$, which is also a maximal torus of ${\rm U}(3,\mathbb{C})$. Note that this subgroup $T$ is Abelian. For $u$, there must exist an $h\in T$ and a $g\in{\rm U}(3,\mathbb{C})$ such that $gug^{-1} = h$. Because $T$ is Abelian, we know that any $t\in T$ commutes with $h$, which implies $tgug^{-1} = gug^{-1}t$. After being left-multiplied by $g^{-1}$ and right-multiplied by $g$, we obtain $g^{-1}tgu = ug^{-1}tg$. Therefore one can see that $\{g^{-1}tg: t\in T \}$ is a subgroup of the centralizer of $\{ u\}$ in ${\rm U}(3,\mathbb{C})$. Furthermore, this group $\{g^{-1}tg: t\in T \}$ is clearly isomorphic to $T$. On the other hand, we know that the center of ${\rm U}(3,\mathbb{C})$ is $\{\begin{pmatrix} \lambda & 0 & 0\\ 0 & \lambda & 0\\ 0 & 0 & \lambda\end{pmatrix}: \lambda\in\mathbb{C}, |\lambda|=1\}$. It is then obvious that $T$ quotient the center of ${\rm U}(3,\mathbb{C})$ still contains many non-identity elements with finite orders, such as $\begin{pmatrix} 1 & 0 & 0\\ 0 & -1 & 0\\ 0 & 0 & -1\end{pmatrix}$ with order $2$ and $\begin{pmatrix} 1 & 0 & 0\\ 0 & i & 0\\ 0 & 0 & -i\end{pmatrix}$ with order $4$. This means $T$ quotient the center of ${\rm U}(3,\mathbb{C})$ is not torsion-free. Thus, $\{g^{-1}tg: t\in T \}$ quotient the center of ${\rm U}(3,\mathbb{C})$ is also not torsion-free. Therefore the centralizer of $\{ u\}$ in ${\rm U}(3,\mathbb{C})$ quotient the center of ${\rm U}(3,\mathbb{C})$ is also not torsion-free.\\
Now we have found a contradiction: the centralizer of the subset $S$ of ${\rm GL}(3,\mathbb{C})$ quotient the center of ${\rm GL}(3,\mathbb{C})$ is torsion-free, but the centralizer of $\{ u\}$ in ${\rm U}(3,\mathbb{C})$ quotient the center of ${\rm U}(3,\mathbb{C})$ is not torsion-free. Finally, we can conclude that there does not exist any isomorphism between ${\rm U}(3,\mathbb{C})$ and ${\rm GL}(3,\mathbb{C})$, i.e., ${\rm U}(3,\mathbb{C})$ and ${\rm GL}(3,\mathbb{C})$ are not isomorphic. 
\end{proof}   
\noindent We would like to remark that the above proof can be easily generalized to $n\geq 3$. In the above proof, only the purely group-theoretic properties of ${\rm U}(3,\mathbb{C})$ and  ${\rm GL}(3,\mathbb{C})$ are used. \textbf{One can now see that there are indeed some intrinsic differences between ${\rm U}(n,\mathbb{C})$ and ${\rm GL}(n,\mathbb{C})$. These subtleties set up warning signs when one is trying to blindly apply conclusions from one to the other.} \\
By the way, although totally unnecessary from the viewpoint of any mathematician, we would like to make some explanation about the meaning of so-called ``up to isomorphism''. This phrase is used in section 2 of \cite{wyz2} immediately after the incorrect assertion that there is an isomorphism between ${\rm U}(n,\mathbb{C})$ and  ${\rm GL}(n,\mathbb{C})$, but seems to be misunderstood by Wang et al. For example, when one says that there is a one-to-one correspondence between a set of groups and another set of groups up to isomorphism, it means that if the group $A$ in one set corresponds to two groups $\mathcal{A}_{1}$ and $\mathcal{A}_{2}$ in the other set, then $\mathcal{A}_{1}$ and $\mathcal{A}_{2}$ are isomorphic. It does not mean that $A$ is isomorphic to $\mathcal{A}_{1}$ or $\mathcal{A}_{2}$. Hopefully our wordy explanation can promote understanding.     \\
From the view of the interplay between Lie groups and algebraic groups, ${\rm U}(n,\mathbb{C})$ and ${\rm GL}(n,\mathbb{C})$ is a commonly mentioned pair of compact Lie group and its complexification. This will be the topic of next subsection.

\subsection{The relation between compact Lie groups and reductive groups via complexification}
From the previous subsection, one should have some feelings about how richness and subtleties can arise from extra structures equipped on abstract groups.\\
 Lie groups and algebraic groups are characterized by different extra structures equipped on the underlying abstract groups. But one can build a bridge between them, such as complexification. A well-known and beautiful result is that via complexification of compact Lie groups there is a one-to-one correspondence between compact Lie groups (up to differentiable isomorphism) and reductive complex algebraic groups (up to polynomial isomorphism) (see for example chapter 5 of \cite{vinberg}). In the previous subsection, we have briefly explained what ``up to isomorphism'' means. Examples of such correspondence are ${\rm U}(n,\mathbb{C}) \rightarrow {\rm GL}(n,\mathbb{C})$ and ${\rm SU}(n,\mathbb{C}) \rightarrow {\rm SL}(n,\mathbb{C})$.\\
 As proved in the previous subsection, compact Lie group ${\rm U}(n,\mathbb{C})$ is not isomorphic to its complexification reductive algebraic group ${\rm GL}(n,\mathbb{C})$. Thus, although with intimate connection, ${\rm U}(n,\mathbb{C})$ and ${\rm GL}(n,\mathbb{C})$ are intrinsically different and one should not expect a statement being true for one to be still true for the other one without whole rigorous reasoning. It is the complexification of a compact Lie group that enjoys the nice properties of algebraic groups, instead of the complex Lie group itself. That is why Bröcker and tom Dieck mentioned in their \cite{tom} about looking at the complexification $G_{\mathbb{C}}$ of a compact Lie group $G$ as an algebraic group.\\
 Since we have mentioned ${\rm SU}(n,\mathbb{C})$ and ${\rm SL}(n,\mathbb{C})$, as a bonus for interested readers, we now give two proofs about the generally non-isomorphic relation between them. Here the phrase ``generally non-isomorphic'' means that ${\rm SU}(n,\mathbb{C})$ and ${\rm SL}(n,\mathbb{C})$ are not isomorphic for almost all possible $n$ except some trivial case such as $n=1$. The first proof is in the context of Lie group for any integer $n\geq 2$, as follows:
 \begin{proof}
 For any integer $n\geq 2$, we know that ${\rm SU}(n,\mathbb{C})$ is compact, but ${\rm SL}(n,\mathbb{C})$ is not compact. Therefore there does not exist any diffeomorphism between ${\rm SU}(n,\mathbb{C})$ and ${\rm SL}(n,\mathbb{C})$ for any integer $n\geq 2$. Then we can see that ${\rm SU}(n,\mathbb{C})$ and ${\rm SL}(n,\mathbb{C})$ are not isomorphic for any $n\geq 2$ in the context of Lie group.
 \end{proof}
\noindent The second proof is in the context of abstract group for $n= 3$, as follows. Again, we would like to remark that here the phrase “in the context of abstract group” means all extra structures equipped on groups, such as topological and geometrical properties, are ignored and only the purely group-theoretic properties are considered in the analysis. We have made this proof as detailed as possible (unnecessarily detailed for most mathematicians) so that for interested readers only basic knowledge of group theory is required. 
\begin{proof}
Suppose ${\rm SU}(3,\mathbb{C})$ and ${\rm SL}(3,\mathbb{C})$ are isomorphic, with one isomorphism denoted by $\sigma:{\rm SL}(3,\mathbb{C}) \rightarrow {\rm SU}(3,\mathbb{C})$. Then for any subset $S'$ of ${\rm SL}(3,\mathbb{C})$, the centralizer of subset $S'$ of ${\rm SL}(3,\mathbb{C})$ is isomorphic to the centralizer of subset $\sigma(S')$ of ${\rm SU}(3,\mathbb{C})$. We choose $S' = \{ \begin{pmatrix} 1 & 1 & 0\\ 0 & 1 & 1\\ 0 & 0 & 1\end{pmatrix}\}\subset {\rm SL}(3,\mathbb{C})$. Then the centralizer of the subset $S'$ of ${\rm SL}(3,\mathbb{C})$ is the subgroup of ${\rm SL}(3,\mathbb{C})$ with elements $\begin{pmatrix} a'_{1} & a'_{2} & a'_{3}\\ b'_{1} & b'_{2} & b'_{3}\\ c'_{1} & c'_{2} & c'_{3}\end{pmatrix}$ satisfying $\begin{pmatrix} a'_{1} & a'_{2} & a'_{3}\\ b'_{1} & b'_{2} & b'_{3}\\ c'_{1} & c'_{2} & c'_{3}\end{pmatrix} \begin{pmatrix} 1 & 1 & 0\\ 0 & 1 & 1\\ 0 & 0 & 1\end{pmatrix} = \begin{pmatrix} 1 & 1 & 0\\ 0 & 1 & 1\\ 0 & 0 & 1\end{pmatrix} \begin{pmatrix} a'_{1} & a'_{2} & a'_{3}\\ b'_{1} & b'_{2} & b'_{3}\\ c'_{1} & c'_{2} & c'_{3}\end{pmatrix}$. It is easy to obtain $b'_{1}=c'_{1}=c'_{2}=0$, $a'_{1} = b'_{2} = c'_{3}$ and $a'_{2} = b'_{3}$. Therefore the centralizer of the subset $S'$ of ${\rm SL}(3,\mathbb{C})$ is $\{ \begin{pmatrix} a'_{1} & a'_{2} & a'_{3}\\ 0 & a'_{1} & a'_{2}\\ 0 & 0 & a'_{1}\end{pmatrix}: a'_{1},a'_{2},a'_{3}\in\mathbb{C}, (a_{1}')^{3} =1 \}$. Since the center of ${\rm SL}(3,\mathbb{C})$ is $\{\begin{pmatrix} \lambda' & 0 & 0\\ 0 & \lambda' & 0\\ 0 & 0 & \lambda'\end{pmatrix}: \lambda'\in\mathbb{C}, (\lambda')^{3} = 1 \}$, one can see that the centralizer of the subset $S'$ of ${\rm SL}(3,\mathbb{C})$ quotient the center of ${\rm SL}(3,\mathbb{C})$ is simply $\{ \begin{pmatrix} 1 & a'_{2} & a'_{3}\\ 0 & 1 & a'_{2}\\ 0 & 0 & 1\end{pmatrix}: a'_{2},a'_{3}\in\mathbb{C} \}$. For a given pair $(a'_{2},a'_{3})\in\mathbb{C}^{2}$, suppose there exists a positive integer $p$ such that $\begin{pmatrix} 1 & a'_{2} & a'_{3}\\ 0 & 1 & a'_{2}\\ 0 & 0 & 1\end{pmatrix}^{p} =  \begin{pmatrix} 1 & 0 & 0\\ 0 & 1 & 0\\ 0 & 0 & 1\end{pmatrix}$. The information retrieved from the $(1,2)$-entry or the $(2,3)$-entry tells us that $a'_{2}   = 0$. With $a'_{2} = 0$ and the information retrieved from the $(1,3)$-entry, we have  $a'_{3} = 0$. Hence the only elements in the centralizer of the subset $S'$ of ${\rm SL}(3,\mathbb{C})$ quotient the center of ${\rm SL}(3,\mathbb{C})$ with finite order is the identity matrix. Therefore by definition the centralizer of the subset $S'$ of ${\rm SL}(3,\mathbb{C})$ quotient the center of ${\rm SL}(3,\mathbb{C})$ is torsion-free.\\
Next we focus on ${\rm SU}(3,\mathbb{C})$. The map $\sigma$ takes $\begin{pmatrix} 1 & 1 & 0\\ 0 & 1 & 1\\ 0 & 0 & 1\end{pmatrix}$ to some element in ${\rm SU}(3,\mathbb{C})$, named $v$ for convenience. We only know some partial information of this element in ${\rm SU}(3,\mathbb{C})$, such as its inifnite order, but not the concrete form. However, we know that any element in ${\rm SU}(3,\mathbb{C})$ is conjugate to some element in the maximal torus $T'=\{\begin{pmatrix} e^{i\theta'_{1}} & 0 & 0\\ 0 & e^{i\theta'_{2}} & 0\\ 0 & 0 & e^{i\theta'_{3}}\end{pmatrix}: \theta'_{1},\theta'_{2},\theta'_{3}\in\mathbb{R}, e^{i( \theta'_{1}+\theta'_{2}+\theta'_{3})}=1\}$. Note that this subgroup $T'$ is Abelian. For $v$, there must exist an $h'\in T'$ and a $g'\in{\rm SU}(3,\mathbb{C})$ such that $g'v(g')^{-1} = h'$. Because $T'$ is Abelian, we know that any $t'\in T'$ commutes with $h'$, which implies $t'g'v(g')^{-1} = g'v(g')^{-1}t'$. After being left-multiplied by $(g')^{-1}$ and right-multiplied by $g'$, we obtain $(g')^{-1}t'g'v = v(g')^{-1}t'g'$. Therefore one can see that $\{(g')^{-1}t'g': t'\in T' \}$ is a subgroup of the centralizer of $\{ v\}$ in ${\rm SU}(3,\mathbb{C})$. Furthermore, this group $\{(g')^{-1}t'g': t'\in T' \}$ is clearly isomorphic to $T'$. On the other hand, we know that the center of ${\rm SU}(3,\mathbb{C})$ is $\{\begin{pmatrix} \lambda' & 0 & 0\\ 0 & \lambda' & 0\\ 0 & 0 & \lambda'\end{pmatrix}: \lambda'\in\mathbb{C}, (\lambda')^{3}=1\}$. It is then obvious that $T'$ quotient the center of ${\rm SU}(3,\mathbb{C})$ still contains many non-identity elements with finite orders, such as previously mentioned $\begin{pmatrix} 1 & 0 & 0\\ 0 & -1 & 0\\ 0 & 0 & -1\end{pmatrix}$ with order $2$ and $\begin{pmatrix} 1 & 0 & 0\\ 0 & i & 0\\ 0 & 0 & -i\end{pmatrix}$ with order $4$. This means $T'$ quotient the center of ${\rm SU}(3,\mathbb{C})$ is not torsion-free. Thus, $\{(g')^{-1}t'g': t'\in T' \}$ quotient the center of ${\rm SU}(3,\mathbb{C})$ is also not torsion-free. Therefore the centralizer of $\{ v\}$ in ${\rm SU}(3,\mathbb{C})$ quotient the center of ${\rm SU}(3,\mathbb{C})$ is also not torsion-free.\\
Now we have found a contradiction: the centralizer of the subset $S'$ of ${\rm SL}(3,\mathbb{C})$ quotient the center of ${\rm SL}(3,\mathbb{C})$ is torsion-free, but the centralizer of $\{ v\}$ in ${\rm SU}(3,\mathbb{C})$ quotient the center of ${\rm SU}(3,\mathbb{C})$ is not torsion-free. Finally, we can conclude that there does not exist any isomorphism between ${\rm SU}(3,\mathbb{C})$ and ${\rm SL}(3,\mathbb{C})$, i.e., ${\rm SU}(3,\mathbb{C})$ and ${\rm SL}(3,\mathbb{C})$ are not isomorphic.
\end{proof}

\subsection{The conditions in Hilbert's finiteness theorem}
In Dieudonné and Carrell's \cite{old}, which is cited by Wang et al. for their understanding of ``reductive groups'', one important subtlety about the sufficient conditions in Hilbert's finiteness theorem is explicitly expressed in the corollary of the theorem in section 1 of chapter 3. Dieudonné and Carrell's corollary is a version of finiteness theorem applicable to compact Lie groups with all finite dimensional linear representations being completely reducible. For rational group action, the sufficient conditions for the relevant ring of invariants being finitely generated over $\mathbb{C}$ consist of \textbf{not only} that every finite dimensional rational representation of the group of our interests (for example ${\rm U}(n,\mathbb{C})$) is completely reducible. Another important condition should also be included to ensure the finiteness, which is stated in section 1 of chapter 3 in \cite{old} as the first assumption for the the group of our interests to satisfy. We restate it here for the convenience of interested readers. Let $R = K[a_{1},...,a_{n}]$ be a finitely generated algebra over an arbitrary field $K$ (here we are interested in the case $K = \mathbb{C}$) and $\Gamma$ be a group of algebra automorphisms of $R$ (here we are interested in the case $\Gamma = {\rm U}(n,\mathbb{C})$). Then the condition needed for the subalgebra $I_{\Gamma}\subset R$ (here we are  interested in the ring of ${\rm U}(n,\mathbb{C})$ invariant of $\{H_{l},H_{\nu},G_{l\nu}^{(k)} \}$) to be finitely generated over $K$ is that \textbf{the orbit under $\Gamma$ of each $f\in R$ is contained in a finite dimensional subspace of $R$ over $K$} \cite{old}. This condition is not a trivial one and cannot be directly absorbed in the condition that every finite dimensional rational representation of the group of our interests (for example ${\rm U}(n,\mathbb{C})$) is completely reducible. One can also find the relevant description in section 1 of Nagata's famous \cite{nagata2}. Needless to say, it is not appropriate to choose $R$ to be the ring of ${\rm U}(n,\mathbb{C})$ invariant of $\{H_{l},H_{\nu},G_{l\nu}^{(k)} \}$ at the beginning, otherwise there will be a circular argument. Therefore, before giving a finitely generated algebra $R$ satisfying this conditions with $\Gamma = {\rm U}(n,\mathbb{C})$ and containing the ring of ${\rm U}(n,\mathbb{C})$ invariant of $\{H_{l},H_{\nu},G_{l\nu}^{(k)} \}$ as a subalgebra, one definitely cannot directly conclude from the complete reducibility of all finite dimensional linear representations of ${\rm U}(n,\mathbb{C})$ that this ring is finitely generated over $\mathbb{C}$.                      \\
On the other hand, when we work in the context of algebraic group, Hilbert's finiteness theorem can be neatly formulated as follows (see for example section 2.2.1 of \cite{derksen} or section 2 of \cite{derksen2}):                
\begin{theorem}
\label{t5}
If $G$ is a linearly reductive group and $V$ is a rational representation, then $K[V]^{G}$ is finitely generated over $K$.
\end{theorem}
\noindent Here $K$ is an algebraically closed field such as $\mathbb{C}$. In the case of $K=\mathbb{C}$, which has characteristic $0$, an algebraic group is linear reductive if and only if it is reductive \cite{nagata}.            \\
\textbf{Therefore, with rational group action, when we say that the (linear) reductiveness of a group can alone imply the finiteness of the generating set of a specific ring of invariants, it is the (linearly) reductive group in the context of algebraic group that we are talking about}. Together with the proofs in the previous section that ${\rm U}(n,\mathbb{C})$ and ${\rm GL}(n,\mathbb{C})$ are not isomorphic as Lie groups for arbitrary positive integer $n$ and not isomorphic in the context of abstract group for integers $n\geq 3$, now one should be able to understand that why we say there is a logical gap in Wang et al.'s \cite{wyz}. We provide one way in \cite{lu} to fill in this logical gap by making use of the isomorphism between the ring of ${\rm U}(n,\mathbb{C})$ invariants of $j$ complex square matrices $\{x_{1},...,x_{j}\}$ and the ring of ${\rm GL}(n,\mathbb{C})$ invariants of $2j$ matrices $\{x_{1},...,x_{j},x_{1}^{\dagger},...,x_{j}^{\dagger}\}$ \cite{procesi}. The finiteness theorem applicable to the ring of ${\rm GL}(n,\mathbb{C})$ invariants of $2j$ complex square matrices $\{x_{1},...,x_{j},x_{1}^{\dagger},...,x_{j}^{\dagger}\}$ has been given and proved in section 3 of \cite{procesi} as his theorem 3.4 (a). With the above reasoning, we can eventually conclude that the ring of ${\rm U}(n,\mathbb{C})$ invariants of $\{H_{l},H_{\nu},G_{l\nu}^{(k)} \}$ is finitely generated over $\mathbb{C}$. And the fact that ${\rm U}(n,\mathbb{C})$ is the maximal compact subgroup of ${\rm GL}(n,\mathbb{C})$ is the reason why the Haar measure of ${\rm U}(n,\mathbb{C})$ is applied in the generalized Molien-Weyl formula for calculating the relevant Hilbert series. \textbf{Note that there may still exist some overlooked gap in the reasoning of our \cite{lu} or this manuscript. We welcome any relevant comments and improvements.}   \\
When stating and applying a mathematical theorem, one should be very careful about the relevant conditions and definitions. A seemingly inconspicuous change may immediately lead to counterexamples. One well-known example closely relevant to our discussion is Nagata's counterexample \cite{nagata3} to the original version of Hilbert's fourteenth problem.                    \\
Since we are talking about Hilbert's finiteness theorem, it is interesting to look back on some relevant history of this theorem starting even before the birth of Hilbert's fourteenth problem. For the development of Hilbert's finiteness theorem, Popov has made a concise summary in his \cite{popov}, which is directly quoted here for the convenience of interested readers: 
\begin{displayquote}
The first fundamental theorem was proved by Gordan in 1868 for invariants of binary forms \cite{gordan} (and in 1870 for systems of binary forms \cite{gordan2}). In 1890 Hilbert gave a general nonconstructive proof for any system of forms in any number of variables \cite{hilbert1}. Although in his version he deals with special representations of GL, the idea of the proof can be carried over directly to the general case. H. Weyl \cite{weyl2} accomplished this for any representation of a complex semisimple group, and Mumford \cite{mum} extended this result to the general case of a reductive algebraic group acting regularly on an affine algebraic variety over a field $k$ of characteristic zero. Finally, through the joint efforts of Nagata \cite{nagata2}, Haboush \cite{haboush}, and Mumford \cite{mum}, the restriction on the characteristic of $k$ was removed (in the case ${\rm char}\ k \neq 0$ Hilbert's approach does not work). In this final form the theorem is now called Hilbert's theorem on invariants.
\end{displayquote}
From the above description one can clearly see the special role of algebraic groups in Hilbert's finiteness theorem.

\section{Summary}
Mathematics is a free land, on which everyone can freely make (logically self-consistent) definitions and name what she/he has defined. Much more important than the names of concepts are the logic and the context within which these concepts are discussed and interplaying. One example closely relevant to our discussion is that Dieudonné and Carrell's definition of ``reductive group'' in their \cite{old} is not totally equivalent to Nagata's definition in \cite{nagata2}. They can still arrive at essentially the same conclusion because logic is universal. Another example is the so-called ``real algebraic group'' used by some mathematicians, which is related to but intrinsically different from what we call algebraic group in algebraic geometry.      \\
We explain in \cite{lu} and more detailedly in this manuscript why there is a considerable logical gap in the argument of \cite{wyz}. There are certainly other paths from ${\rm U}(2,\mathbb{C})$ and ${\rm U}(3,\mathbb{C})$, as compact Lie groups with all finite dimensional linear representations being completely reducible, to the finiteness of the generating sets of unitary invariants of $\{H_{l},H_{\nu},G_{l\nu}^{(k)} \}$. One of these paths is discussed in our \cite{lu}, which makes use of the isomorphism between the ring of ${\rm U}(n,\mathbb{C})$ invariants of $j$ complex square matrices $\{x_{1},...,x_{j}\}$ and the ring of ${\rm GL}(n,\mathbb{C})$ invariants of $2j$ matrices $\{x_{1},...,x_{j},x_{1}^{\dagger},...,x_{j}^{\dagger}\}$. We believe Wang et al. and other interested readers can also construct their own (straightforward or twisting) paths if they want.\\
Neither in \cite{lu} nor in this manuscript we deny the correctness of the calculation results in Wang et al.'s \cite{wyz}. But such correctness can be justified only by a whole logical chain, instead of vague arguments or wishful thinking. The motive of this manuscript and \cite{lu} is to point out why there is a gap and how to deal with it. \textbf{After all, a gap is a gap before it is filled in.}

\end{document}